\documentclass{article}

\usepackage{PRIMEarxiv}

\usepackage[utf8]{inputenc} 
\usepackage[T1]{fontenc}    
\usepackage{hyperref}       
\usepackage{url}            
\usepackage{booktabs}       
\usepackage{amsfonts}       
\usepackage{nicefrac}       
\usepackage{microtype}      
\usepackage{lipsum}
\usepackage{fancyhdr}       
\usepackage{graphicx}       
\usepackage{subcaption}      
\graphicspath{{media/}}     
\usepackage{multicol}
\usepackage{multirow}
\usepackage{adjustbox}

\title{Mutual Learning for Finetuning Click-Through Rate Prediction Models}

\author{
  İbrahim Can Yılmaz \\
  AI Enablement \\
  Huawei Türkiye R\&D Center \\
  Istanbul, Turkey\\
  \texttt{ibrahim.can.yilmaz2@huawei-partners.com} \\
   \And
  Said Aldemir \\
  AI Enablement \\
  Huawei Türkiye R\&D Center \\
  Istanbul, Turkey\\
  \texttt{said.aldemir1@huawei.com} \\
}

\begin{document}
\maketitle

\begin{abstract}
Click-Through Rate (CTR) prediction has become an essential task in digital industries, such as digital advertising or online shopping. Many deep learning-based methods have been implemented and have become state-of-the-art models in the domain. To further improve the performance of CTR models, Knowledge Distillation based approaches have been widely used. However, most of the current CTR prediction models do not have much complex architectures, so it’s hard to call one of them ‘cumbersome’ and the other one ‘tiny’. On the other hand, the performance gap is also not very large between complex and simple models. So, distilling knowledge from one model to the other could not be worth the effort. Under these considerations, Mutual Learning could be a better approach, since all the models could be improved mutually. In this paper, we showed how useful the mutual learning algorithm could be when it is between equals. In our experiments on the Criteo and Avazu datasets, the mutual learning algorithm improved the performance of the model by up to 0.66\% relative improvement.
\end{abstract}


\section{Introduction}
The prediction of the click through rate (CTR) is a significant task in online advertising and recommendation systems. Accurate prediction of user behavior on a specific ad can lead to efficient ranking of potential ad candidates and achieving more clicks on displayed ads. This can directly affect the revenue of the advertiser and the publisher. Also, displaying the relevant ads to each user can enhance the user satisfaction on the service that recommendation model operates.

Many deep learning-based CTR models achieved significant success in recommending ads. The models on the literature mainly based on feature interactions and user behavior modelling. Factorization machine (FM)-based models have been proposed to derive the second-order feature interactions. In order to capture higher order feature interactions, multi-layer perceptron (MLP) based models have been also introduced.

Large-scale deep learning models bring model complexity, significant hardware requirements, and high training cost despite the success in computer vision and natural language processing tasks. Also, deploying these complex models in real-life applications is challenging due to the limited resources. To overcome this problem, several techniques are proposed in the literature. Knowledge distillation is a very popular method widely studied in the literature to compress the size of deep learning models.

Knowledge distillation concept is firstly introduced by Hinton et al. \cite{hinton2015distilling}. In this technique, the knowledge learned by a large deep learning model is transferred to the relatively small network. This structure consists of a complex teacher model and a small student model. The idea is to train the student model with the teacher network so that the student network mimics the complex network and achieves better or competitive performance relative to the teacher network.

The concept of mutual learning is an extension of the knowledge distillation proposed by Zhang et al. \cite{zhang2017deep}. Unlike the teacher-student structure in knowledge distillation, student models learn collaboratively while the training. The advantages of this technique are to not depend on formerly trained teacher network and to enable student models to teach themselves in the training process.

Several studies have used the concept of knowledge distillation to improve the prediction accuracy of CTR models in the recommendation literature. Zhu et al. proposed an ensemble distillation architecture consists of multiple teacher networks with a gating mechanism and a student network \cite{zhu2023ensemble}. In this study, gating mechanism is used to adaptive, sample-wise learning from multiple teacher models. 

In another study, the graph neural network (GNN) based bridge module is proposed for model training with knowledge distillation in the CTR prediction task \cite{DengBrigeBasedKD}. The knowledge learned by the teacher network is first transferred to the bridge module, then distilled into the student network. A GNN-based bridge module is used to catch the important feature interactions.

Furthermore, Guan et al. utilized the conventional knowledge distillation framework in order to reduce the complexity of their proposed model that is deep interaction compressed network (DICN) \cite{GUAN2023110704}. In addition, they stated that the integration of the knowledge distillation framework into the model training improves the generalization ability of the model, so the prediction performance increased with respect to the baseline.

Here we present a detailed investigation of mutual learning approaches for CTR models with extensive experiments and satisfying scores that prove that knowledge distillation can improve model scores in addition to its capability on model compression. In the following sections of this publication, we will first describe the methodologies employed in this study, followed by a discussion of the experiments under four major questions.

\section{Methods}
\subsection{Deep CTR Models}
Deep-learning based CTR prediction models have shown remarkable success in various web services, including online advertising, news recommendation, and web search. These models primarily leverage the capabilities of deep neural networks to capture high-order feature interactions \cite{ChengWideDeep, LianXDeepFM}. Additionally, other CTR models like Deep Interest Network \cite{zhou2018din} and Deep Interest Evolution Network \cite{zhou2018dien} take into account users' historical behavior to forecast future behavior for specific ad requests. In this study, we utilize the following deep CTR models to illustrate the effectiveness of the deep mutual learning concept.

\textbf{DeepFM} \cite{guo2017deepfm}: DeepFM is the one of the fundamental networks in the CTR prediction task. It combines the FM and MLP modules in the network architecture. 

\textbf{DCN} \cite{WangDCN} : Deep \& Cross Network jointly learn the low-order and higher-order feature interactions with the proposed architecture consisting of fully-connected feed-forward neural network and multilayer cross network. The output is calculated by the sigmoid layer after concatenated the outputs of these two networks.

\textbf{PNN} \cite{QuPNN}: Product-based Neural Network proposes two types of the product networks that are inner product-based and outer product-based neural networks in order to learn high-order latent patterns. In the architecture, product layer is followed by the MLP layer and the prediction is calculated with the sigmoid layer. 

\textbf{FiBiNET} \cite{Huang_2019}:  The Feature Importance and Bilinear Interaction Network dynamically captures feature importance and detailed feature interactions through the use of the Squeeze-Excitation network (SENET) mechanism and the Bilinear-Interaction layer.

\subsection{Knowledge Distillation}
Knowledge distillation framework consists of training a complex teacher network and distilling the knowledge learned by the teacher model into the light student network. In the Figure 1, the training scheme for knowledge distillation is showed. Firstly, heavy teacher model trained on the CTR dataset and the training loss is defined between the model prediction and ground-truth labels. After that, the student model is trained with two loss functions that are distillation loss resulted from the soft label predictions of the teacher and student models and also the student loss between the student model predictions and hard labels.

\begin{figure}
  \centering
  \resizebox{0.5\linewidth}{!}{\includegraphics{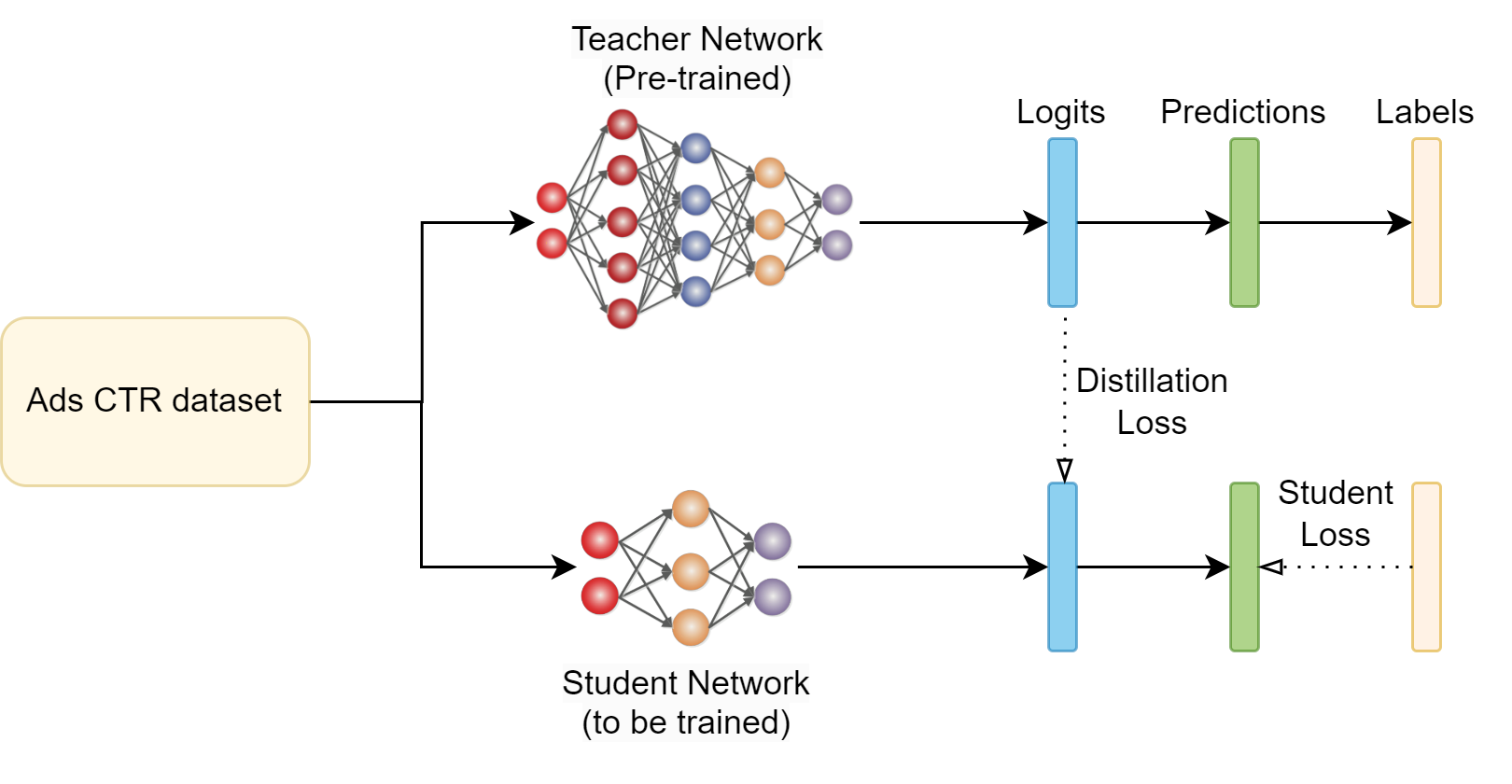}}
  \caption{Architecture of Knowledge Distillation algorithm.}
  \label{fig:fig1}
\end{figure}

\subsection{Deep Mutual Learning}
In the deep mutual learning technique, the student models are trained at the same time, unlike knowledge distillation. This technique eliminated the dependency of the distillation method on a powerful teacher network that is pretrained before distilling the knowledge to the student network. In the scheme, the lightweight student models learn the problem simultaneously with two different losses. The first one is the supervised learning loss, and the other one is the mutual loss between the student model predictions. By this technique, it is stated that the student networks learn better than the networks that are trained in a conventional distillation framework with a large teacher network.

\subsection{Proposed Algorithm}
The CTR prediction task is commonly regarded as a binary classification problem because its goal is to determine whether or not the item is clicked. Its output is a single floating-point number representing the probability of the item being clicked. Therefore, we used Mean Squared Error (MSE) loss as the mutual loss. We take a weighted sum of the mutual loss and supervised learning loss, namely Binary Cross Entropy (BCE) loss, as the final loss for each model. The architecture of the algorithm is shown in the Figure 2. MSE loss and BCE loss can be formulated as follows.

\begin{equation}
    L_{MSE}(p_1, p_2) = (p_1 - p_2)^2
\end{equation}
\begin{equation}
    L_{BCE}(y, p) = -(y \cdot log(p) + (1 - y) \cdot log(1 - p))
\end{equation}

where \(p, p_1, p_2\) are model predictions representing click probabilities and \(y\) is the label indicating if the item is clicked or not. Given N networks \((\theta_1, \theta_2, \cdots, \theta_N)\), we can calculate the loss function for each network trained with mutual learning algorithm as follows.

\begin{equation}
    L_{\theta_n} = L_{BCE_n} + \frac{1}{N - 1} \sum_{i=1,i \neq n}^N L_{MSE}(p_i, p_n)
\end{equation}

\begin{figure}
  \centering
  \resizebox{0.5\linewidth}{!}{\includegraphics{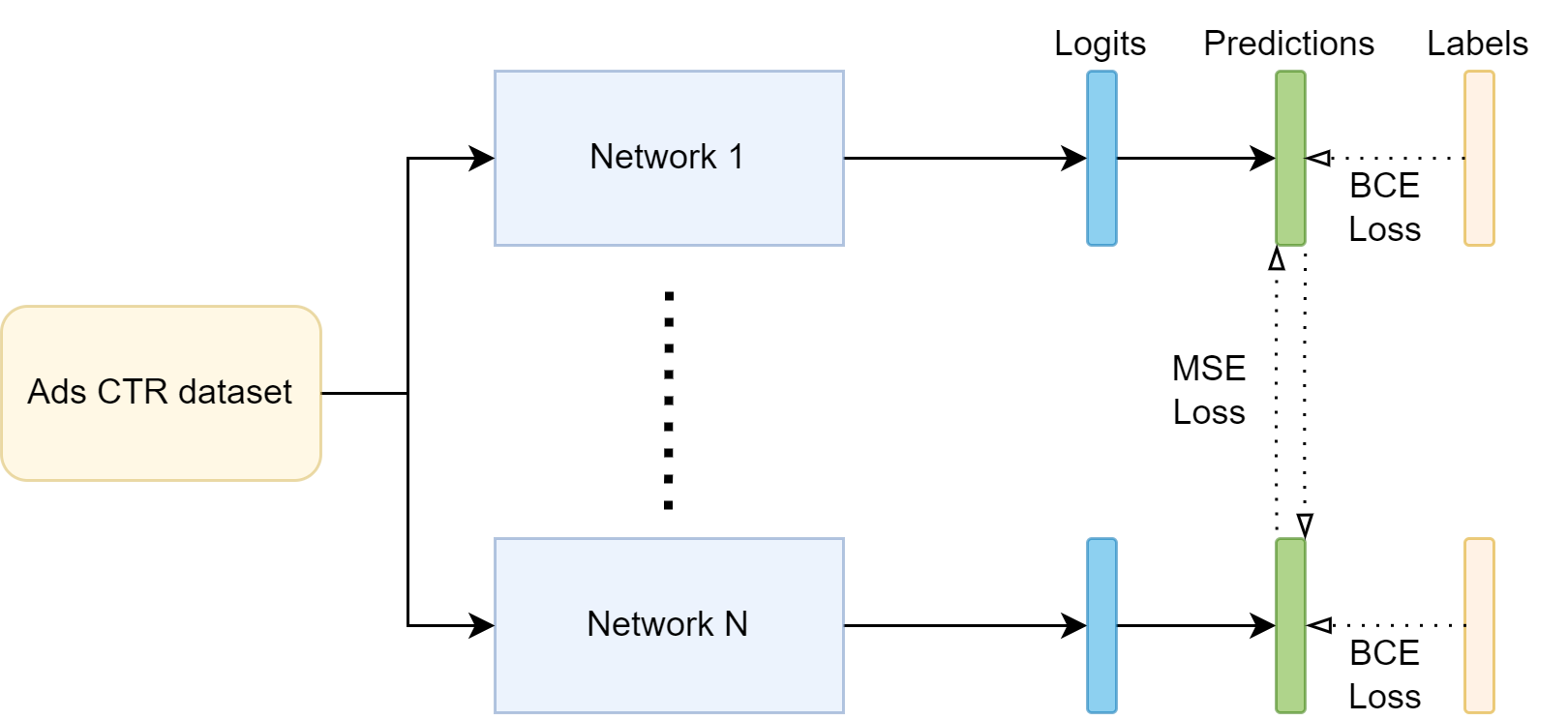}}
  \caption{Architecture of proposed algorithm.}
  \label{fig:fig2}
\end{figure}

\section{EXPERIMENTS \& RESULTS}
We conducted extensive experiments to test the effectiveness of mutual learning for click-through rate models. In this section we’re trying to answer following questions:

\begin{itemize}
    \item \textbf{RQ1:} How successful are models trained with the mutual learning strategy in comparison with the same models trained independently?
    \item \textbf{RQ2:} How does pre-training effect the quality of mutual learning approach?
    \item \textbf{RQ3:} Is it more beneficial to train different models mutually than training multiple instances of the same model?
    \item \textbf{RQ4:} How important is the number of co-trained models to improve the model performances?
\end{itemize}

In the following subsections, we will first describe the experimental setup, then go over the experiments in detail, and finally interpret the results.

\subsection{Experimental Setup}
\subsubsection{Datasets}
We tested our approach on two datasets widely used in the click-through rate prediction task. Both datasets were split in train-dev-test sets by 80\%, 10\%, 10\% ratios, respectively.

\textbf{Criteo:} The Criteo dataset is commonly used in the evaluation of CTR model performances. It comprises click logs with more than 45M instances. Criteo dataset includes 26 anonymus categorical features and 13 continuous numerical features.

\textbf{Avazu:} The Avazu dataset consists of  40M click logs which is ordered chronologically. There are 24 fields including a unique id for each instance, click information and 22 categorical features.

\subsubsection{Evaluation Metrics}
AUC (Area Under ROC curve) and relative improvement of AUC (RelaImp) were adopted for the evaluation of algorithms.

\textbf{AUC:} It’s widely used for imbalanced classification tasks, espacially. This is because it doesn’t depend on a classification threshold. It basically measures whether a positive sample ranked higher than the negatives. It is formulated as:

\begin{equation}
    AUC(f) = \frac{\sum_{s^n \in S^n} \sum_{s^p \in S^p} 1[f(s^n) < f(s^p)]}{|S^n| \cdot |S^p|}
\end{equation}

where \(f\) is the predictor, \(s^n\) is a negative sample in \(S^n\), the set of negative examples, and \(s^p\) is a positive sample in \(S^p\), the set of positive examples. \(1[f(s^n )<f(s^p )]\) denotes an indicator that returns 1 if the probability that the predictor returns for the positive sample \(s^p\) is higher than the one for the negative sample \(s^n\), otherwise it returns 0.

\textbf{RelaImp:} Measures the improvement relatively against a baseline score for any metric. Here, we will be using it to evaluate how mutual learning improves the AUC scores compared to independently trained models. Formulation of RelaImp is as following:

\begin{equation}
    RelaImp = \frac{AUC_{test} - 0.5}{AUC_{baseline} - 0.5} -1
\end{equation}

\subsubsection{Test Models}
We chose four of most common and foundational models of click-through rate prediction task: FiBiNET, DCN, PNN, and DeepFM. These models have been widely used for click-through rate prediction and have become inspiration for newer state-of-the-art models.

\subsubsection{Hyperparameters and Implementation Details}
For a fair comparison, all common hyperparameters were kept the same in all models. Datasets were split into mini-batches with the size of 4000 sample. Embedding size was set to 16. L2 regularization coefficient for the embeddings was set to \(1 \times 10^{-5}\). Hidden layer sizes of MLP modules of each model was set to [400, 200, 100]. L2 regularization coefficient for MLP layers was set to \(1 \times 10^{-7}\). The learning rate was set to \(1 \times 10^{-3}\) for independent trainings and mutual trainings from scratch and \(7 \times 10^{-4}\) for mutual trainings on pretrained models, which were the best values for each scenario. The learning rate was scheduled during training with an exponential decay function that reduces the learning rate to one tenth in three epochs.

\subsection{RQ1: Comparison with regular training}
To evaluate the effectiveness of the mutual learning strategy, we initially compared its outcomes with those of independently trained instances of the same models. Each of the four models underwent training four times, resulting in four AUC scores per model. Subsequently, four randomly initialized instances of the same model were generated and trained collaboratively using the proposed algorithm. Consequently, for each model, we obtained four AUC scores from the mutual learning algorithm as well.

Table 1 shows the results for both models trained independently and models trained with mutual learning from scratch. Apparently, mutual learning does not improve results when it is applied from scratch. Although the standard deviation in the AUC scores of the models trained with mutual learning is very low, they are still slightly worse than the models trained independently on average.

\begin{table*}
    \caption{Independent Training vs. Mutual Learning}
    \centering
    \begin{adjustbox}{max width=\linewidth}
    \label{table:results}
    \renewcommand{\arraystretch}{1.2}
    \begin{tabular}{|l|l|cccc|cccc|}
    \hline
        \multicolumn{2}{|c|}{Independent Training vs.} & \multicolumn{8}{c|}{Datasets} \\ \cline{3-10}
        \multicolumn{2}{|c|}{Mutual Learning} & \multicolumn{4}{c|}{Avazu} & \multicolumn{4}{c|}{Criteo} \\ \cline{3-10}
        \hline
        \multirow{3}{*}{FiBiNET} & Independent Runs & 0.79743 & 0.79720 & 0.79691 & 0.79648 & 0.81311 & 0.81310 & 0.81321 & 0.81308  \\
        ~ & Mutual from scratch & 0.79701 & 0.79695 & 0.79698 & 0.79706 & 0.81307 & 0.81314 & 0.81309 & 0.81306  \\ 
        ~ & Mutual with pretraining & 0.79837 & 0.79834 & 0.79842 & 0.79831 & 0.81425 & 0.81420 & 0.81426 & 0.81425  \\ \hline
        \multirow{3}{*}{DCN} & Independent Runs & 0.79425 & 0.79394 & 0.79383 & 0.79390 & 0.81251 & 0.81249 & 0.81248 & 0.81243  \\
        ~ & Mutual from scratch & 0.79385 & 0.79392 & 0.79407 & 0.79398 & 0.81243 & 0.81242 & 0.81245 & 0.81239  \\
        ~ & Mutual with pretraining & 0.79562 & 0.79565 & 0.79578 & 0.79560 & 0.81330 & 0.81329 & 0.81327 & 0.81328  \\ \hline
        \multirow{3}{*}{PNN} & Independent Runs & 0.79595 & 0.79586 & 0.79577 & 0.79622 & 0.81242 & 0.81235 & 0.81231 & 0.81220  \\
        ~ & Mutual from scratch & 0.79587 & 0.79591 & 0.79583 & 0.79586 & 0.81228 & 0.81237 & 0.81234 & 0.81230  \\
        ~ & Mutual with pretraining & 0.79704 & 0.79706 & 0.79689 & 0.79704 & 0.81343 & 0.81336 & 0.81337 & 0.81330  \\ \hline
        \multirow{3}{*}{DeepFM} & Independent Runs & 0.79607 & 0.79623 & 0.79600 & 0.79584 & 0.81206 & 0.81208 & 0.81206 & 0.81208  \\
        ~ & Mutual from scratch & 0.79598 & 0.79593 & 0.79601 & 0.79597 & 0.81199 & 0.81203 & 0.81207 & 0.81201  \\
        ~ & Mutual with pretraining & 0.79700 & 0.79672 & 0.79680 & 0.79670 & 0.81295 & 0.81295 & 0.81301 & 0.81296  \\ \hline
    \end{tabular}
    \renewcommand{\arraystretch}{1}
    \end{adjustbox}
\end{table*}

\subsection{RQ2: Mutual Learning as finetuning}
Since mutual learning didn’t give the improvement we expected, we decided to use the independently trained models as pretrained models for mutual learning. Training the pretrained models with mutual learning strategy for one more epoch made 0.30-0.66\% relative improvement on Avazu dataset and 0.25-0.37\% relative improvement on Criteo dataset, as shown in the Table 1. After one epoch, performance started to degrade for all models.

It must be discussed why mutual learning after pretraining made such significant improvements while mutual learning from scratch did not make any at all. There is no obvious conclusion here, but our interpretation is as follows:

\begin{itemize}
    \item Every single one of the model instances learns in a unique way, depending on the initialization and randomness. Hence, each instance learns something different when they are trained separately. Distilling their unique knowledge to each other creates a combination of experts. This is why mutual learning works with pretraining.
    \item Training from scratch mutually prevents instances from learning uniquely, because it constrains what instances learn distilling the knowledge from the very beginning of training. Thence, instances learn the same information during the training. This is why mutual learning doesn't work from scratch.
\end{itemize}

\subsection{RQ3: Same model instances vs. different models}
The main reason for questioning whether training different models or multiple instances of the same model is more beneficial for mutual learning is the question of whether the models learn some information specific to the model itself—and can this information be transferred between models? To answer this question, we take the independently trained weights with the best score for each model as pretrained models for mutual training. As a result, the performance of the models increased significantly, as expected. However, as Table 2 shows, the improvement is lower than the same-model training (RQ2) for the model with the best score, while the results are even better for models with lower scores.

In light of these results, we can assume that the models do not learn information specific to their model class. To be clear, FiBiNET model instances don’t learn such information that only FiBiNET models can learn, but DCN, PNN, or DeepFM cannot. In fact, by evaluating Table 2 together with Table 1, we can conclude that a model benefits from mutual learning better when its accompanying models have higher pretrained scores.

\subsection{RQ4: Does the number of co-trained models matter?}
We know that mutual learning improves the performance of pretrained models now; however, what number of co-trained models gives the best results is still an open question. To investigate this, we took four models pretrained from the first step and dropped one with the worst AUC score at a time. Hence, we tested mutual learning with 4, 3, and 2 pretrained models, respectively. We applied this test to FiBiNET on the Avazu dataset and to DCN on the Criteo dataset, for the sake of diversity.

According to the results in Tables 3 and 4, performance decreased marginally when we reduced the number of instances to three, but decreased significantly more when we reduced it to two. Here, we can see that even the worst-pretrained model contributes to the performance improvement. This could mean that models learn something unique in every single training, as long as the initialization is random.

To determine the upper bound, we increased the number of instances to five, which, on average, almost did not make any impact on the results. Thus, we conclude that the performance reaches a plateau when the number of instances is four.

\begin{table}[!ht]
    \caption{Comparing AUC scores of Mutual Learning with same model instances vs. different model instances}
    \centering
    \begin{adjustbox}{max width=0.5\linewidth}
    \renewcommand{\arraystretch}{1.5}
    \begin{tabular}{|l|l|l|l|l|l|}
    \hline
        \multicolumn{2}{|c|}{Same Model vs.} & \multicolumn{4}{c|}{Models} \\ \cline{3-6}
        \multicolumn{2}{|c|}{Different Models} &  FiBiNET & DCN & PNN & DeepFM  \\ \hline
        \multirow{3}{*}{Avazu} & Pretrained & 0.79743 & 0.79425 & 0.79622 & 0.79623  \\
        ~ & DML of same models & \textbf{0.79837} & 0.79562 & 0.79704 & 0.79672  \\
        ~ & DML of diff. models & 0.79800 & \textbf{0.79622} & \textbf{.79715} & \textbf{0.79677}  \\ \hline
        \multirow{3}{*}{Criteo} & Pretrained & 0.81321 & 0.81251 & 0.81242 & 0.81208  \\
        ~ & DML of same models & \textbf{0.81426} & 0.81330 & 0.81343 & 0.81295  \\
        ~ & DML of diff. models & 0.81415 & \textbf{0.81336} & \textbf{0.81355} & \textbf{0.81320}  \\ \hline
    \end{tabular}
    \renewcommand{\arraystretch}{1}
    \end{adjustbox}
\end{table}

\begin{table*}[!ht]
    \caption{Comparing AUC scores according to number of co-trained models in mutual learning – avazu – fibinet}
    \centering
    \begin{adjustbox}{max width=\linewidth}
    \renewcommand{\arraystretch}{1.5}
    \begin{tabular}{|l|ccccc|ccccc|}
    \hline
        \multicolumn{1}{|c|}{Number of} & \multicolumn{10}{|c|}{Avazu} \\ \cline{2-11}
        \multicolumn{1}{|c|}{Co-trained Models} & \multicolumn{5}{c|}{Pretrained models} & \multicolumn{5}{c|}{Finetuning with Mutual Learning}   \\ \hline
        \multirow{4}{*}{FiBiNET} & 0.79743 & 0.79720 & 0.79691 & 0.79701 & 0.79648 & 0.79838 & 0.79835 & 0.79839 & 0.79833 & 0.79831  \\
        ~ & 0.79743 & 0.79720 & 0.79691 & 0.79701 & - & 0.79837 & 0.79834 & 0.79842 & 0.79835 & -  \\
        ~ & 0.79743 & 0.79720 & 0.79691 & - & - & 0.79829 & 0.79837 & 0.79833 & - & -  \\
        ~ & 0.79743 & 0.79720 & - & - & - & 0.79813 & 0.79811 & - & - & -  \\ \hline
    \end{tabular}
    \renewcommand{\arraystretch}{1}
    \end{adjustbox}
\end{table*}

\begin{table*}[!ht]
    \caption{Comparing AUC scores according to number of co-trained models in mutual learning – criteo – dcn}
    \centering
    \begin{adjustbox}{max width=\linewidth}
    \renewcommand{\arraystretch}{1.5}
    \begin{tabular}{|l|ccccc|ccccc|}
    \hline
        \multicolumn{1}{|c|}{Number of} & \multicolumn{10}{|c|}{Criteo} \\ \cline{2-11}
        \multicolumn{1}{|c|}{Co-trained Models} & \multicolumn{5}{c|}{Pretrained models} & \multicolumn{5}{c|}{Finetuning with Mutual Learning}   \\ \hline
        \multirow{4}{*}{DCN} & 0.81251 & 0.81249 & 0.81248 & 0.81247 & 0.81243 & 0.81290 & 0.81330 & 0.81328 & 0.81327 & 0.81328  \\
        ~ & 0.81251 & 0.81249 & 0.81248 & 0.81247 & - & 0.81330 & 0.81329 & 0.81327 & 0.81328 & -  \\
        ~ & 0.81251 & 0.81249 & 0.81248 & - & - & 0.81325 & 0.81326 & 0.81321 & - & -  \\
        ~ & 0.81251 & 0.81249 & - & - & - & 0.81323 & 0.81325 & - & - & -  \\ \hline
    \end{tabular}
    \renewcommand{\arraystretch}{1}
    \end{adjustbox}
\end{table*}

\section{Conclusion}
Motivated by the achievements of knowledge distillation in a variety of machine learning domains, we proposed a mutual learning algorithm to fine-tune click-through rate prediction models. Because most of the CTR models do not have such complex or heavy model structures, we considered mutual learning to be the key to transferring knowledge between models. We investigated the effectiveness of the mutual learning algorithm through four key questions. First, we compared the performance of the models trained mutually from scratch with the ones trained independently. We found out that it does not improve the results; it even decreases a little, even though its results are more robust and have a lower standard deviation. However, when we tried mutual training with the pretrained models, it significantly improved their performances. Therefore, we concluded that each model instance learns unique information and applying mutual learning from scratch constrains that feature. Instead of training multiple instances of the same model mutually, we trained different models together. It showed that performance decreased for the model with the highest initial score, but improved for the rest. We interpreted this as meaning that the models can have different learning capabilities, yet they don't learn anything model-specific. However, we have learned that the better the accompanying models, the more a model will benefit from mutual learning. Finally, we investigated the optimal number of models to train together.

In conclusion, we investigated different aspects of mutual learning algorithm on click-through rate prediction task. We showed that mutual learning can make a significant improvement when it’s used for finetuning. The fact that the results are consistent for both datasets and all models is an indication that our observations are solid and stable.

\bibliographystyle{unsrt}  
\bibliography{main}

\end{document}